\documentclass[10pt]{article} 

\usepackage[sorting=none,citestyle=numeric-comp]{biblatex}
\usepackage[margin=1in]{geometry}
\usepackage{fancyhdr}
\fancypagestyle{plain}{%
\fancyhf{} 
\fancyfoot[R]{-\,\fontsize{8pt}{8pt}\selectfont\thepage\,-} 

}
\RequirePackage{amsthm,amsmath,amsfonts,amssymb}
\RequirePackage{graphicx, colortbl}
\usepackage{arydshln}

\newcommand{\totem}{\textsc{Tot}\textsc{Em}\xspace}

\usepackage{enumitem}

\usepackage{dsfont}

\usepackage{graphicx}
\usepackage{sidecap}
\usepackage{xspace} 
\usepackage{mathtools}
\usepackage{multirow}
\usepackage{hyperref}

\usepackage{thmtools,thm-restate}

\usepackage{letltxmacro}
\LetLtxMacro{\originaleqref}{\eqref}
\renewcommand{\eqref}{Eq.~\originaleqref}

\newcommand{\multidistro}{\text{mult}\xspace}

\newcommand{\equ}[1]{\begin{gather} #1 \end{gather}}
\newcommand{\sfrac}[2]{\mbox{$\frac{#1}{#2}$}}
\newcommand{\abs}[1]{\left\vert#1\right\vert} 
\newcommand{\abss}[1]{\vert#1\vert} 
\newcommand{\quads}[1]{\quad #1 \quad}
\newcommand{\qand}{\quad \text{and} \quad}

\newcommand{\prob}[1]{\mathfrak{#1}}

\makeatletter
\newcommand*\bigcdot{\mathpalette\bigcdot@{.7}}
\newcommand*\bigcdot@[2]{\mathbin{\vcenter{\hbox{\scalebox{#2}{$\m@th#1\bullet$}}}}}
\makeatother

\DeclarePairedDelimiterX\infdivx[2]{(}{)}{
  #1\;\delimsize\|\;#2
}
\DeclarePairedDelimiterX\braket[2]{\langle}{\rangle}{
  #1\;\vert\;#2
}
\DeclarePairedDelimiterX\braketO[3]{\langle}{\rangle}{
  #1\;\vert#2\vert\;#3
}

\newcommand{\infdiv}[2]{D\hspace{0.1mm}\infdivx{#1}{#2}}

\usepackage{subfig}

\usepackage[colorinlistoftodos,prependcaption,textsize=small]{todonotes}

\definecolor{beaublue}{rgb}{0.74, 0.83, 0.9}
\definecolor{mintcream}{rgb}{0.96, 1.0, 0.98}
\newcommand{\tablecolor}{beaublue}
\definecolor{cadmiumgreen}{rgb}{0.0, 0.42, 0.24}

\usetikzlibrary{tikzmark,decorations.pathreplacing,calligraphy,positioning,calc,arrows.meta,shapes.geometric,shapes.arrows}
\tikzstyle{vertex} = [draw=black, shape=circle,node distance=55pt]
\tikzstyle{phantomvertex} = [fill=white,shape=circle,node distance=30pt]
\tikzstyle{edge} = [fill,opacity=.3,fill opacity=.3,line cap=rect, line join=round, line width=50pt]
\tikzstyle{hyperedge} = [fill,opacity=.3,fill opacity=.3,line cap=round, line join=round, line width=30pt]

\begin{document}

\thispagestyle{empty}

\begin{flushright}
\phantom{Version: \today}
\\
\end{flushright}
\vskip .2 cm
\subsection*{}
\begin{center}
{\Large {\bf On the estimation and interpretation \\[1ex] of effect size metrics
} }
\\[0pt]

\bigskip
\bigskip {\large
{\bf Orestis Loukas}\footnote{E-mail: orestis.loukas@uni-marburg.de},\,
{\bf Ho Ryun Chung}\footnote{E-mail: ho.chung@uni-marburg.de}\bigskip}\\[0pt]
\vspace{0.23cm}
{\it Institute for Medical Bioinformatics and Biostatistics\\
Philipps-Universität Marburg\\
Hans-Meerwein-Straße 6, 35032 Germany}

\bigskip
\end{center}

\begin{abstract}
\noindent
Effect size estimates are thought to capture the collective, two-way response to an intervention or exposure in a three-way problem among the intervention/exposure, various confounders and the outcome. 
For meaningful causal inference from the estimated effect size, the joint distribution of observed confounders must be identical across all intervention/exposure groups. 
However, real-world observational studies and even randomized clinical trials often lack such structural symmetry. 
To address this issue, various methods have been proposed and widely utilized. Recently, elementary combinatorics and information theory have motivated a consistent way to completely eliminate observed confounding in any given study.
In this work, we leverage these new techniques to evaluate  conventional methods based on their ability to (a) consistently differentiate between collective and individual responses to intervention/exposure and (b) establish the desired structural parity for sensible effect size estimation.
Our findings reveal that a straightforward application of logistic regression 
homogenizes the three-way stratified analysis, but fails to restore structural symmetry leaving in particular the two-way effect size estimate unadjusted. 
Conversely, the Mantel–Haenszel estimator struggles to separate three-way effects from the two-way effect of intervention/exposure, leading to inconsistencies in interpreting pooled estimates as two-way risk metrics.
\end{abstract}


\section{Introduction}

Randomized clinical trials (\textsc{rct}) aim at detecting and quantifying the causal response to an intervention or exposure such as some  treatment or risk factor. 
From the onset, we distinguish between the response of individual subjects and the collective response of a cohort of individuals to concrete external conditions.
Since there usually exist many variables --\,so called confounders\,-- that influence the individual response at the level of each subject, the study design generically pertains to isolating the collective  from individual effect of intervention/exposure on the outcome to facilitate a fair comparison across treatments or risk factors. 

The initial step toward causal inference involves estimating the two-way effect size associated solely with the intervention/exposure without reference to individual subjects.
%
Contrary to a rationale often encountered in the literature~\cite{fuller2019confounding}, all study groups must exhibit the same distribution over observed profiles, i.e.\ manifestations of confounders. In other words, it is not enough to balance e.g.\ \texttt{gender} and \texttt{age} independently. It is the joint distribution of those two factors which needs to be replicated over all intervention/exposure groups. 
%
Formally, the ideal study structure is realized as a two-way symmetry --\,called in~\cite{loukas2023eliminating} structural \textsc{p}arity\,-- between observed confounders and intervention/exposure groups:
\equ{
\label{eq:StructuralParity:Symmetry}
\text{group A}~ \rightarrow ~\text{group B} : \quad
\text{Pr}(\text{confounders}\vert \text{group A}) = \text{Pr}(\text{confounders}\vert \text{group B})
}
Since the conditional distribution of confounders remains invariant under study groups, it empirically coincides with the global distribution  $\text{Pr}(\text{confounders})$. %
In such  structurally homogeneous setting,  it  suffices for the purpose of causal inference to estimate the collective effect of intervention/exposure on the outcome variable without referring to particular subject profiles, i.e.\ without worrying about confounders, even when the latter significantly contribute to the outcome. 

In practice, finite sample sizes 
prohibit exact randomization resulting in disparate distributions of confounder profiles across study groups. 
Due to structural dis\textsc{p}arity (also refereed to as structural heterogeneity in this context), confounding to one or more confounder variables could arise. 
The potentially confounder-induced bias complicates an estimation of the collective effect size of intervention/exposure directly from the study data, which in turn obfuscates causal inference.
A naive, so called crude, effect size estimate would generically capture the two-way response to intervention alongside --\,through structural dis\textsc{p}arities across intervention/exposure groups\,-- the influence of individual profiles on the outcome. 
Since the cost of redesigning and conducting an \textsc{rct} can be great, combined with further technical aspects and ethical concerns, the importance of consistently dealing with confounding at the level of statistical analysis becomes higher.
Beyond \textsc{rct}, observational studies deliver important information about dependencies among features as well as the design of future experimental studies. Due to their historic character, such studies cannot be modified evading randomization.  Therefore, structural dis\textsc{p}arities and potentially associated confounding effects are unavoidable in observational studies.

In~\cite{loukas2023eliminating}, the authors of present paper successfully applied combinatorial and information-theoretic techniques --\,first used in an analogous context in~\cite{loukas2023demographic}\,-- in order to achieve a hypothetical scenario which eliminates  structural heterogeneity from historic \textsc{rct}s and observational studies. In that way, the problem of observed confounding was consistently addressed with the guarantees of well-established concepts from information theory. Given the observed structurally dis\textsc{p}arate study, the new approach (in short the \textsc{pr}-projection) uniquely re-estimates the collective effect size of intervention/exposure under the symmetric situation \eqref{eq:StructuralParity:Symmetry}. 
Similar to fair-aware machine learning, the symmetric alternative of biomedical studies treats all intervention/exposure groups in a ``fair'' way by specifically prescribing one global confounder distribution to each group. 

The goal of this paper is to compare and benchmark against more conventional methods  which try to address the adverse effects of confounding in clinical trials and observational studies. 
%
Relying exclusively on information-theory, we recapitulate the minimal ingredients that are epistemologically needed to achieve confounding-free effect size estimates that reflect as much of the original trends in the data as mathematically possible. 
Using the optimal three-way contingency table with complete structural \textsc{p}arity, we 
benchmark model-centric (logistic regression, logit) and heuristic (Mantel–Haenszel, \textsc{mh}) methods in order to draw conclusions on their applicability and interpretation.  In this exploration, we fully uncover in Section~\ref{ssc:Logit} the meaning of logit and pinpoint that its intended use does not coincide with the aims of de-confounding a study, at all. 
At the same time, we raise serious questions about the content and interpretability of \textsc{mh} estimators as two-way metrics in Section~\ref{ssc:Benchmark}, while even demonstrating their internal inconsistencies in the concrete setting of Section~\ref{sc:SymmetricExample}.
To our knowledge, this leaves the suggested method of the \textsc{pr}-projection as the only formally viable and epistemologically consistent option in confounding-free analysis.

\section{Methods}

Confounding-free analysis necessitates a meticulous adjustment of the data from a conducted study to retrospectively ``correct'' for some observed heterogeneity in the study design.
Therefore, we are interested in hypothetical studies with an identical structure across intervention/exposure groups, beyond the received data. 
Instead of assuming models fundamentally in clash with the data or postulating some adjusted effect size estimate to later try to uncover its meaning, we deductively proceed from givens to hypothesis to ultimately reach the optimal solution. 
In the spirit of \textsc{tot}al \textsc{em}piricism (\totem~\cite{loukas2023total}), deconfounding comprises a 
high-precision modification of dependencies among features with a clear physical meaning. In contrast to machine learning and artificial intelligence, there are neither black boxes nor random explorations in this thorough investigation.

First, we separate the features naturally used to characterize subjects in a dataset into three parties: the response $Y$ with possible outcomes $\mathcal D_Y$, attributes $\boldsymbol X$ defining intervention/exposure groups $\mathcal D_{\boldsymbol X}$  and confounders $\boldsymbol S$ describing subject profiles $\mathcal D_{\boldsymbol S}$ in the study. 
A three-way contingency table 
consists of summary statistics of subjects' microdata that specify a count for each manifestation in the Cartesian product $\mathcal D=\mathcal D_Y\times \mathcal D_{\boldsymbol X} \times D_{\boldsymbol S}$. An expected contingency table can be described by a distribution $\prob p\equiv\prob p_{Y,\boldsymbol X,\boldsymbol S}$ on the simplex over $\mathcal D$, which quantifies our perception of individual response to the intervention/exposure. 
A special distribution is the empirical $\prob f$ whose probabilities equal the relative cell counts from the actual study. 

By definition, a three-way contingency table contains the information on all conceivable dependencies among the recorded features
with any additional input a priori regarded as unintentional bias.
For the purposes of the present paper, we can concentrate 
on marginal distributions (in short marginals) among $n$ features to unambiguously quantify their $n$-point dependency.
%
Below, we optimally derive the  expectation of a three-way contingency table $\prob q$ whose two-way marginal satisfies the desired structural \textsc{p}arity across study groups. 
Epistemologically, the expected contingency table $\prob q$ is necessary to rigorously
deduce any hypothetical effect size estimate in the confounding-free scenario without relying on realistically unfulfilled limits and model assumptions.
Only through $\prob q$, we can unequivocally relate contingency tables between $Y$ and $\boldsymbol X$ stratified by $\boldsymbol S$ with the effect size estimate of intervention/exposure computed from the two-way contingency table between $Y$ and $\boldsymbol X$.

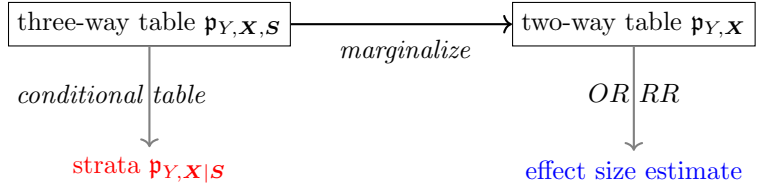
\begin{figure}[!t]
    \centering
    \begin{tikzpicture}[scale=.92, remember picture ] 
    \node (rect) at (0,0) [draw] (threeway) {three-way table $\prob p_{Y,\boldsymbol X,\boldsymbol S}$};
    \node (rect) at (7,0) [draw] (twoway) {two-way table $\prob p_{Y,\boldsymbol X}$};
    \draw[->, thick] (threeway) -- (twoway);
    \node at (3.7,-0.4)  (mar) {\textit{marginalize}};
    \node (rect) at (0,-2.1) [red] (strata) {strata  $\prob p_{Y,\boldsymbol X\vert\boldsymbol S}$};
    \node (rect) at (7,-2.1) [blue] (effectsize) {effect size estimate};
    \draw[->, thick, gray] (threeway) -- (strata);
    \draw[->, thick, gray] (twoway) -- (effectsize);
    \node at (-0.55,-1) [rotate=0] (OR1) {\textit{conditional table}};
    \node at (7.,-1)  (OR2) {\textit{$OR$ $RR$}};
    \end{tikzpicture}
    \caption{The top-down approach to confounder analysis. Given a three-way contingency table, we marginalize designated confounders to obtain a two-way contingency table between response and intervention/exposure. From the latter, one can estimate the effect size instead of using the stratified analysis at the three-factor level.}
    \label{fig:Top-Down}
\end{figure}
The epistemologically sensible way to compute risk metrics such as odds ratios ($OR$) and relative risks ($RR$) in any hypothetical scenario 
is summarized by the flow Diagram~\ref{fig:Top-Down}. Suppose $\prob p$ had been determined by some optimality criterium, its stratified conditional distribution
$\prob p_{Y,\boldsymbol X\vert \boldsymbol S}$
immediately follows so that we can compute stratified odds,
\equ{
Odds_{\prob p} (\texttt{event}\vert \boldsymbol x,\boldsymbol s) 
= \frac{p_{Y \vert \boldsymbol X, \boldsymbol S}(\texttt{event}\vert \boldsymbol x,\boldsymbol s)}{p_{Y\vert \boldsymbol X,\boldsymbol S}(\texttt{no\,\,event} \vert \boldsymbol x,\boldsymbol s)} 
= 
\frac{p_{Y,\boldsymbol X\vert \boldsymbol S}(\texttt{event}, \boldsymbol x\vert\boldsymbol s)}{p_{Y,\boldsymbol X\vert \boldsymbol S}(\texttt{no\,\,event}, \boldsymbol x\vert\boldsymbol s)} 
}
as ordinary three-way odds. Only after marginalizing $\boldsymbol S$ to obtain a two-way contingency table deducing conditional distribution $\prob p_{Y\vert \boldsymbol X}$, we could proceed with the estimation of the effect size collectively assigned to intervention/exposure $\boldsymbol x$ quantified e.g.\ based on
\equ{
Odds_{\prob p}(\texttt{event}\vert\boldsymbol x) = \frac{p_{Y\vert \boldsymbol X}(\texttt{event}\vert\boldsymbol x)}{p_{Y\vert \boldsymbol X}(\texttt{no\,\,event}\vert\boldsymbol x)}
~.
}
From there, it is straight-forward to compare study groups by calculating the
\equ{
OR_{\prob p} = \frac{Odds_{\prob p}(\texttt{event}\vert\boldsymbol x)}{Odds_{\prob p}(\texttt{event}\vert\boldsymbol x_0)}~.
}
As an alternative metric for the effect size of intervention/exposure, the relative risk is often estimated:
\equ{
RR_{\prob p} = \frac{\prob p(\texttt{event}\vert \boldsymbol x)}{\prob p(\texttt{event}\vert \boldsymbol x_0)}
}

Therefore, an expectation about the three-way contingency table consistently and non-trivially relates the effect size contingent on individual profiles with the collective effect size estimate used in the derivation of hypothetical risk metrics. 
Next, we need a mechanism to systematically incorporate given information into our confounding-free hypothesis and deduce such expected contingency table. For that, we can turn to sampling processes.
If all we have is the data $\prob f$ of the actual study with $N$ participants, then sampling would  be naturally performed from $\prob f$. 
In population statistics, elementary combinatorics establish the sampling distributions
\equ{
\multidistro(N\prob p;\prob f) = N!\prod\limits_{y,\boldsymbol x,\boldsymbol s}\frac{(f(y,\boldsymbol x,\boldsymbol s))^{Np(y,\boldsymbol x,\boldsymbol s)}}{(Np(y,\boldsymbol x,\boldsymbol s))!} 
\qand
\text{hypergeoemtric}(N\prob p; M\prob f) = 
\frac{\prod\limits_{y,\boldsymbol x,\boldsymbol s}
\binom{M f(y,\boldsymbol x,\boldsymbol s)}
{Np(y,\boldsymbol x,\boldsymbol s)}
}
{
\binom{M}{N}
}
~,
}
where $n!=1\cdots n$ denotes the factorial and $\binom{n}{m}=\frac{n!}{m!(n-m)!}$ the binomial coefficient.
Assuming a population of size $M \gg N^2 \gg 1$ the large $M$-$N$ expansion  
\equ{
\log \text{hypergeoemtric}(N\prob p; M\prob f)
=
-N \infdiv{\prob p}{\prob f} - \frac{\abss{D}-1}{2}\log \left(2\pi N\right) - \sfrac12\sum_{Y\in\mathcal D_Y} \sum_{\boldsymbol x\in\mathcal D_{\boldsymbol X}}\sum_{\boldsymbol s\in\mathcal D_{\boldsymbol S}}\log p(y,\boldsymbol x,\boldsymbol s)
+ \ldots 
}
reveals that sampling is universally governed by the information divergence 
\equ{
\label{eq:KLmetric}
\infdiv{\prob p}{\prob f} = \sum_{Y\in\mathcal D_Y} \sum_{\boldsymbol x\in\mathcal D_{\boldsymbol X}}\sum_{\boldsymbol s\in\mathcal D_{\boldsymbol S}} p(y,\boldsymbol x,\boldsymbol s) \log \frac{p(y,\boldsymbol x,\boldsymbol s)}{f(y,\boldsymbol x,\boldsymbol s)}
}
of some hypothetical study $\prob p$ from the original study. 
Intuitively, the larger the divergence, the less likely to sample such hypothetical study under $\prob f$.
 
Clearly, sampling from  $\prob f$ would trivially make the original study  most likely under itself. If we hypothesized certain alternative conditions however to lift the observed heterogeneity by restoring structural symmetry \eqref{eq:StructuralParity:Symmetry}, $\prob f$ would not generically fulfill them (otherwise there would have been no observed confounding to begin with). 
This automatically raises the question regarding those hypothetical studies that now become most likely, when sampling from the original $\prob f$ which still represents our source of knowledge about what has actually happened.
%
As long as our \textsc{pr} hypothesis
is self-consistent and $N$ suffieciently large, there exists~\cite{csiszar1975divergence} one hypothetical study $\prob q$ in the structurally symmetric scenario where
\equ{
\label{eq:PRconditions}
\text{structural \textsc{p}artiy:}\quad  \prob q_{\boldsymbol X,\boldsymbol S} \overset{!}{=}\prob f_{\boldsymbol X}\cdot \prob f_{\boldsymbol S}
\qand
\text{confounder \textsc{r}ealism}:\quad \prob q_{Y,\boldsymbol S} \overset{!}{=} \prob f_{Y,\boldsymbol S}
}
apply, which exhibits minimal information divergence from the original study and hence becomes most likely under the actual data. 
Luckily for realistic datasets, this optimal\footnote{With an increasing sample size, a significant portion of the likelihood to sample some hypothetical confounding-free study from $\prob f$  concentrates~\cite{Rosenkrantz1989} over studies with decreasing information divergence from the optimal study $\prob q$.} (in the sense of maximum likelihood~\cite{loukas2023total}) study can be efficiently obtained in a fully model-agnostic way using \textsc{ipf}~\cite{loukas2022categorical}, Newton-Raphson~\cite{loukas2022entropy} or any combination thereof.

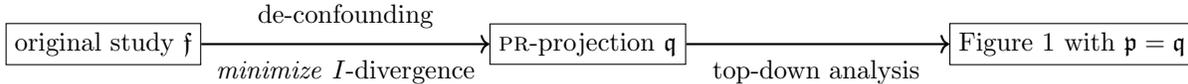
\begin{figure}[!t]
    \centering
    \begin{tikzpicture}[scale=.92, remember picture ] 
    \node (rect) at (0,0) [draw] (f) {original study $\prob f$};
    \node (rect) at (7,0) [draw] (q) {\textsc{pr}-projection $\prob q$};
    \draw[->, thick] (f) -- (q);
    \node at (3.5,0.4)  {de-confounding};
    \node at (3.5,-0.4)  {\textit{minimize} $I$-divergence};
    \node (rect) at (14,0) [draw] (fig1) {Figure~\ref{fig:Top-Down} with $\prob p=\prob q$};
    \node at (10.3,-0.4)  {top-down analysis};
    \draw[->, thick] (q) -- (fig1);
    \end{tikzpicture}
    \caption{The data-centric top-down approach to confounder analysis.}
    \label{fig:data-centric}
\end{figure}
Hence, we obtain through elementary combinatorics at large sample (and population) scales an emerging information theoretic description that maps the original data to one hypothetical study, thus deducing a unique effect size estimate in the confounding-free scenario. The applicability of this empirical data-centric approach is limited neither by the cardinality of $\abs{\mathcal D_{\boldsymbol X}}$ and $\abs{\mathcal D_{\boldsymbol S}}$ nor by the presence of higher $n$-point dependencies. 
In principle, we could choose any expected study $\prob p$ on the simplex over $\mathcal D$ that enjoys structural \textsc{p}arity to calculate two-way effect size metrics as in Figure~\ref{fig:Top-Down}. 
Minimizing \eqref{eq:KLmetric} under the \textsc{pr} hypothesis \eqref{eq:PRconditions} guarantees~\cite{jaynes1968prior,csiszar1991least} that we are not injecting any unintentional bias by an arbitrary choice of $\prob p$.
Model assumptions generically suffer from the latter form of bias:
Even when model-centric confounder analysis follows the top-down approach of flow diagram~\ref{fig:Top-Down}, there is generically no guarantee that unintentional bias 
were injected into the data in the process of de-biasing w.r.t.\ observed confounding.

\section{The association of smoking and lung cancer}

\begin{table}[!t]
    \centering
   \renewcommand{\arraystretch}{1.1}
    \centering
    \begin{tabular}{ll|r}
         \texttt{occupation} & \texttt{age} & $\frac{Odds_{\prob p}(\texttt{case}\vert\texttt{1+\,\,pack\,\,daily}, \texttt{occupation}, \texttt{age})}{Odds_{\prob p}(\texttt{case}\vert \texttt{non-smokers}, \texttt{occupation}, \texttt{age})}$\\
        \hline &&\\[-2.5ex]
        \texttt{housewives} & \texttt{45-54} & 9.60\\
        \rowcolor{\tablecolor}
         & \texttt{45-54} & 9.00\\
         \rowcolor{\tablecolor}
        & \texttt{55-64} & 5.75\\
        \rowcolor{\tablecolor}
        \multirow{-3}{*}{\texttt{white-collar workers}} & \texttt{over 65} & 0\\
        & \texttt{45-54} & 48.00\\
        \multirow{-2}{*}{\texttt{other}} & \texttt{55-64} & 0\\
    \end{tabular} 
    \caption{Three-way $OR$ for lung cancer due to exposure stratified by confounder profiles that can be defined (i.e.\ are finite) from the original study ($\prob  p=\prob f$). on the 
    These are reproduced by the $\textsc{pr}$-projection $\prob p=\prob q$.}
    \label{tb:SmokeLung:StratifiedOR}
\end{table}

In the observational study considered by~\cite{10.1093/jnci/22.4.719}, the effect of smoking (with two exposure groups \texttt{1+\,\,pack\,\,daily} vs.\ \texttt{non-smokers}) on undifferentiated pulmonary carcinoma (\texttt{case}) is investigated. 
The outcome is confounded by (at least)  two variables, the \texttt{occupation} (with three levels) and \texttt{age} group (with four levels) of studied subjects. The three-way table of the study is given on page 20 in~\cite{10.1093/jnci/22.4.719}.

Without making any model assumptions, our starting point should be always the original study described by empirical distribution $\prob f$, which exhibits 
the so-called crude effect size estimate  due to exposure quantified e.g.\ by 
\equ{
\label{eq:Smoking:CrudeEffectSize}
OR_{\prob f} = \frac{Odds_{\prob f}(\texttt{case}\vert \texttt{1+\,\,pack\,\,daily})}{Odds_{\prob f}(\texttt{case}\vert \texttt{non-smokers})} \approx 7.10
~.
}
According to Graph~\ref{graph:PRprojection},
the \textsc{pr} approach derives  
the closest to $\prob f$ 
three-way  contingency table described by joint distribution $\prob q$ that is expected under \eqref{eq:PRconditions}.
In particular, structural \textsc{p}arity between exposure groups
implies conditional independence across all confounder manifestations (combinations of \texttt{occupation} with \texttt{age}):
\equ{
q_{\boldsymbol S\vert X} (\texttt{occupation},\texttt{age}\vert \texttt{non-smokers}) = 
q_{\boldsymbol S\vert X} (\texttt{occupation},\texttt{age}\vert \texttt{1+\,\,pack\,\,daily})
=
f_{\boldsymbol S} (\texttt{occupation},\texttt{age})
~.
} 
The unique solution of the \textsc{pr} optimization problem fully reproduces the empirical stratification by designated confounders in Table~\ref{tb:SmokeLung:StratifiedOR}.
In this top-down approach, it is straight-forward to then calculate the expected effect size estimate,
\equ{
\label{eq:Smokers:PR_Intervention_OR}
OR_{\prob q} = \frac{Odds_{\prob q}(\texttt{case}\vert \texttt{1+\,\,pack\,\,daily})}{Odds_{\prob q}(\texttt{case}\vert \texttt{non-smokers})} \approx 8.61~,
}
of the confounding-free scenario. This hypothetical scenario  adheres to the confounder \textsc{r}ealism (yellow hyperedge in diagram~\ref{graph:PRprojection}) of 
\equ{
\label{eq:Smokers:ConfounderRealism}
Odds_{\prob f}(\texttt{case}\vert \texttt{occupation},\texttt{age})~,
}
which we take to give a fair estimate of population statistics. If one suspects that this is not the case, either an external proxy for the response-confounder association must be provided to the \textsc{pr} optimization problem or the data should not be used at all.

As a technical side remark, some confounder profiles are not observed in both exposure groups, due to sampling zeros (vanishing cell counts in the original contingency table). Retaining all 12 confounder manifestations warrants a form of regularization, which  reduces  in probability space to pseudo-counts, see e.g.\ Appendix A in~\cite{loukas2023total}. The quoted effect size estimate remains stable under any arbitrarily small regularization corresponding to pseudo-counts approaching machine-precision, even when certain three-way risk metrics become arbitrarily large.
Small sample size is generically a problem that reduces statistical power which no model assumption can solve  without introducing unintentional bias.

\paragraph{The original ansatz.}
In contrast to the advocated logic of flow diagram~\ref{fig:Top-Down}, the authors of~\cite{10.1093/jnci/22.4.719} introduced  a pooled effect size estimate 
\equ{
\label{eq:Smoking:Pooled_OR}
\widehat{OR} = 
\frac{\sum_{\boldsymbol s} f(\texttt{case},\texttt{1+\,\,pack\,\,daily},\boldsymbol s)\cdot f(\texttt{no\,\,case}, \texttt{non-smokers},\boldsymbol s)/f_{\boldsymbol S}(\boldsymbol s)}
{\sum_{\boldsymbol s} f(\texttt{case},\texttt{non-smokers},\boldsymbol s)\cdot f(\texttt{no\,\,case}, \texttt{1+\,\,pack\,\,daily},\boldsymbol s)/f_{\boldsymbol S}(\boldsymbol s)}
\approx 10.68~,
}
which became known as the Mantel–Haenszel estimator forming the basis for the Cochran-Mantel–Haenszel statistics. In addition, they list further suggestions, which come close, but never numerically coincide with \eqref{eq:Smokers:PR_Intervention_OR}. Despite lacking a clear epistemological completion to a three-way contingency table, the estimate \eqref{eq:Smoking:Pooled_OR} is perceived to capture some of the characteristics of the confounding-free scenario with structural \textsc{p}arity.
In Section~\ref{ssc:Benchmark} and~\ref{sc:SymmetricExample}, we investigate in a bottom-up manner what the \textsc{mh} estimator does not capture trying to actually find out what it captures.

\usetikzlibrary{shapes.misc}
\tikzset{cross/.style={cross out, draw=black, minimum size=2*(#1-\pgflinewidth), inner sep=0pt, outer sep=0pt}, cross/.default={1pt}} 

\pgfdeclarelayer{bg}    
\pgfsetlayers{bg,main}  

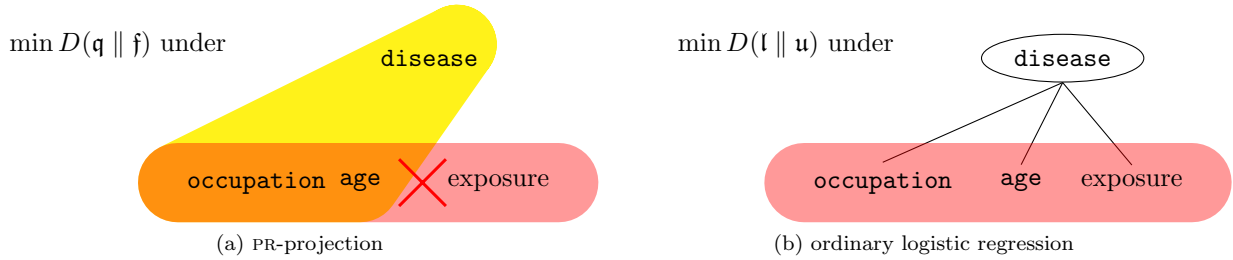
\begin{figure}[t]
\subfloat[\centering \textsc{pr}-projection]{
\begin{tikzpicture}[scale=.92, remember picture] 
\node at (-4.5, 0.2) {$\min\infdiv{\prob q}{\prob f}$ under};
\node at (0,0) (Y) {\texttt{disease}};
\node at (-2.5,-1.8)  (S1) {\texttt{occupation}};
\node at (-1.,-1.8)  (S2) {\texttt{age}};
\node at (1,-1.8)  (X) {\text{exposure}};
\draw (-0.1,-1.8) node[cross=10pt,red,line width=0.4mm,] {};
\begin{pgfonlayer}{bg} 
\draw[hyperedge, opacity=0.9, yellow] (0.4,0.2)  -- (S1.west)  -- (S2.center) -- (0.4,0.2);
\draw[hyperedge, opacity=0.4, red] (S1.west)  -- (X.east)  ;
\end{pgfonlayer}
\end{tikzpicture}\label{graph:PRprojection}
}
\subfloat[\centering ordinary logistic regression]{
\quad\quad
\begin{tikzpicture}[scale=.92, remember picture ] 
\node at (-4., 0.2) {$\min\infdiv{\prob l}{\prob u}$ under};
\node at (0,0) [draw, ellipse] (Y) {\texttt{disease}};
\node at (-2.6,-1.8)  (S1) {\texttt{occupation}};
\node at (-0.6,-1.8)  (S2) {\texttt{age}};
\node at (1,-1.8)  (X) {\text{exposure}};
\begin{pgfonlayer}{bg} 
\draw (Y.south) -- (X.north);
\draw (Y.south) -- (S1.north);
\draw (Y.south) -- (S2.north);
\draw[hyperedge, opacity=0.4, red] (S1.west)  -- (X.east)  ;
\end{pgfonlayer}
\end{tikzpicture}\label{graph:logit}
}
\caption{Derivation of the three-way contingency table by minimizing the $I$-divergence from either (a) empirical  or (b) uniform distribution under two- and three-feature marginal conditions signified by edges and hyperedges, respectively. The \textsc{pr}-projection follows after decoupling \texttt{age} and \texttt{occupation} from \texttt{exposure}.}
\end{figure}

\subsection{Logistic regression}
\label{ssc:Logit}

A model often employed~\cite{agresti2007introduction} in effect size estimation is logistic regression (logit) with joint distribution $\prob l$.
As noted in~\cite{loukas2023total}, logit can be fully understood on the simplex over $\mathcal D$ as a distribution of Maximum Entropy avoiding awkward parametrizations or one-hot encoding. 
According to Graph~\ref{graph:logit}, logit stays as close as possible to the three-way contingency table with equal cell counts (uniform distribution $\prob u$) while learning  
pairwise the original  marginal distribution of \texttt{disease} with each of the predictors:
\equ{
\label{eq:Smoking:logit:PairwiseConstraints}
\prob l_{Y,X} \overset{!}{=} \prob f_{Y,X}
\qand
\prob l_{Y,S_i} \overset{!}{=} \prob f_{Y,S_i} \quads{,}i=1,2~.
}
Partially learning $Y-\boldsymbol S$ results in e.g.\ $Odds_{\prob p}(\texttt{case}\vert\texttt{housewives},\texttt{under 45})$ dropping to $0.08$ for $\prob p=\prob l$ from the empirical $0.29$ for $\prob p=\prob f$.
Moreover, the expected prediction of the trained logit on the train data itself necessarily adheres to  
the combined (generically disparate) distribution of $X=\texttt{exposure}$ with $ S_1=\texttt{occupation}$ and $S_2=\texttt{age}$:
\equ{
\label{eq:Smoking:logit:PredictorConstraint}
\prob l_{X,\boldsymbol S} {=} \prob f_{X,\boldsymbol S}
~.
}

By model design, the logit produces an expectation $\prob l$ for the thee-way contingency table which has regularized and homogenized  (due to the uniform ansatz) the stratified odds ratios 
setting them equal to
\equ{
\label{eq:Smoking:Logit:Stratified_OR}
\frac{Odds_{\prob l}(\texttt{case}\vert \texttt{1+\,\,pack\,\,daily},\texttt{occupation}, \texttt{age})}{Odds_{\prob l}(\texttt{case}\vert\texttt{non-smokers}, \texttt{occupation}, \texttt{age})\phantom{..}} \approx 10.02~.
}
Although this property might seem appealing,  
our initial aim was neither to remove associations\footnote{$n$-point associations with $n\geq3$ can realistically become significant, also in biomedical settings~\cite{psf2022005028}.} pertaining confounder \textsc{r}ealism (specifically the $Y-\texttt{occupation}-\texttt{age}$ three-point effect) nor to homogenize the  
potentially significant three-way effects of the original stratified analysis.
Instead, we want to minimally
adjust the effect size of exposure \eqref{eq:Smoking:CrudeEffectSize} after  eliminating structural dis\textsc{p}arity. 
By construction  however, the logit identically reproduces both two-way effects contained in \eqref{eq:Smoking:logit:PairwiseConstraints} and \eqref{eq:Smoking:logit:PredictorConstraint} from the original study. 
In particular, this retains the two-way effect size estimate \eqref{eq:Smoking:CrudeEffectSize} of the original study 
meaning that the  hypothetical study of logit is still plagued by the observed confounding.

In total, a straight-forward application of logistic regression (a) removes potentially significant three-way effects, (b) leaves the generically disparate distribution of confounders with intervention/exposure groups unaltered and (c) does not re-estimate the effect size of intervention/exposure  which also remains non-adjusted. 
The arising situation, which was  first exemplified in Section~3.1 of~\cite{loukas2023eliminating} at the large sample size limit,
is to be contrasted with the \textsc{pr} projection in Graph~\ref{graph:PRprojection}. Having exemplified the shortcomings of standard logit, we focus on the \textsc{mh} estimator in the following sections.

As an extension of the conventional logit, one can think of applying the logistic conditional distribution $\prob l_{Y\vert X,\boldsymbol S}$ on a hypothetical dataset enjoying complete \textsc{p}arity (hence also conditional independence) between $x=\texttt{1+\,\,pack\,\,daily}$ and $x=\texttt{non-smokers}$: 
\equ{
\label{eq:Smoking:Logit_Parity}
\ell'(\texttt{case}, x,\texttt{occupation}, \texttt{age}) =  \ell_{Y\vert X,\boldsymbol S}(\texttt{case}\vert x, \texttt{occupation}, \texttt{age}) \cdot f_X(x) \cdot f_{\boldsymbol S} (\texttt{occupation}, \texttt{age})~.
}
Such \textsc{p}arity-preserving logistic scenario would raise the crude effect size estimate to approximately 9.45, while maintaining the homogenized three-way odds ratio in \eqref{eq:Smoking:Logit:Stratified_OR}.
We note that the covariant formulation in probability space easily allows for combining conventional modeling with the developed information-theoretic framework. 
However, 
\eqref{eq:Smoking:Logit_Parity} moves away from confounder \textsc{r}ealism, $\infdiv{\prob f_{Y,\boldsymbol S}}{\prob l'_{Y,\boldsymbol S}}\approx 0.012$ against $\infdiv{\prob f_{Y,\boldsymbol S}}{\prob l_{Y,\boldsymbol S}}\approx 0.009$ and $\infdiv{\prob f_{Y,\boldsymbol S}}{\prob q_{Y,\boldsymbol S}}\approx0$ (up to regularization). This property should be generically regarded  as data-incompatible, since 
it proposes unjustified alterations to statistics and strata, primarily based on the presumed ``beauty'' of the sigmoid function.

\subsection{Benchmark under structural variations}
\label{ssc:Benchmark}

\begin{figure}[t]
    \centering
    \raisebox{0.5\height}{
\begin{tikzpicture}[scale=.92, remember picture ] 
\draw (0,0) circle [radius=0.3] node (Y) {$Y$};
\coordinate (s) at (-1.5,-2);
\draw (s) circle [radius=0.3] node (S) {$\boldsymbol S$};
\coordinate (x) at (1.5,-2);
\draw (x) circle [radius=0.3] node (X) {$X$};
\draw (-1.1,-0.8) node[rotate=54] {\textsc{r}ealism};
\draw (0,-2.6) node[red] {structural dis\textsc{p}arity};
\draw (0,-3) node[red] {(controlled by $\delta$)};
\draw (1.2,-0.7) node[rotate=-54,blue] {exposure};
\draw[-] (S) -- (Y);
\draw[-] (S) -- (X);
\draw[-,dashed,blue] (X) -- (Y);
\end{tikzpicture}
}
\quad
\includegraphics[scale=0.8,clip,trim=0mm 5mm 0mm 0mm]{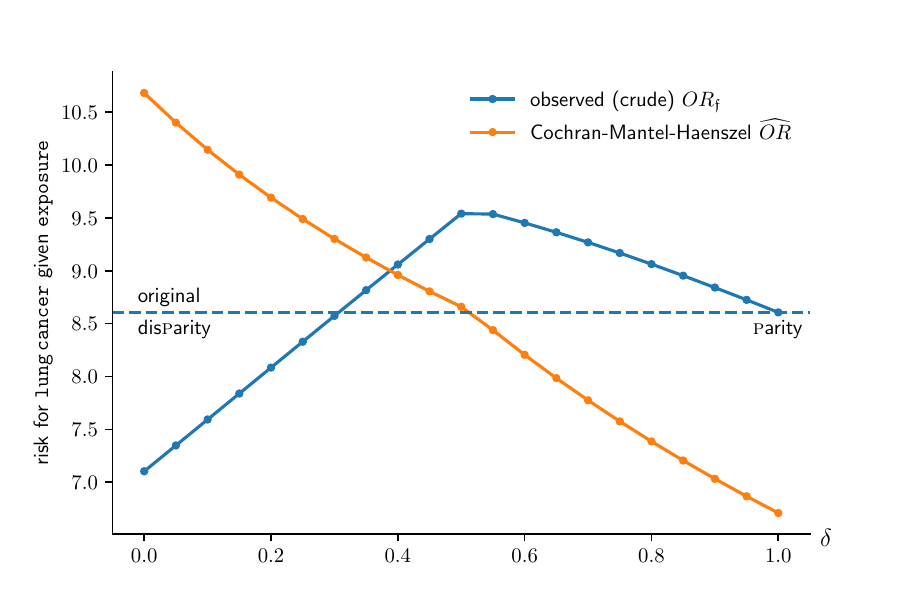}
\caption{Left: Graph for simulated scenarios with varying structural dis\textsc{p}arity based on the original study. By construction, all simulated studies correspond to the same confounding-free estimate (readily deduced under structural \textsc{p}arity $\delta=1$) exhibiting the same stratified Table~\ref{tb:SmokeLung:StratifiedOR}. Right: Observed and \textsc{mh} estimated ${OR}$ pooled under the simulated structural (dis)\textsc{p}arity.}
    \label{fig:simulations}
\end{figure}
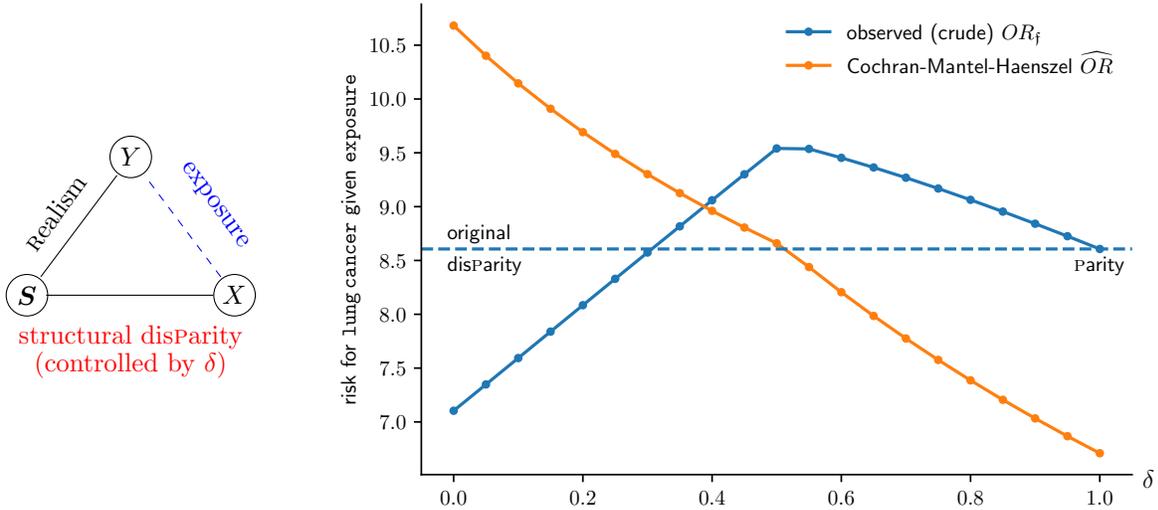
In the previous Section, we found a discrepancy among the \textsc{pr} estimate for the two-way effect size in \eqref{eq:Smokers:PR_Intervention_OR} and the \textsc{mh} pooled estimator \eqref{eq:Smoking:Pooled_OR}. This comes as no surprise after~\cite{loukas2023eliminating}, which uncovered systematic discrepancies between the \textsc{mh} and \textsc{pr} effect size estimates in various historic examples. In this Section, we shall try to clarify the situation by simulating studies of the same biological plausibility (same dependency of the outcome on confounders) but varying level of structural dis\textsc{p}arity. 

Based on the previously studied association of lung cancer with smoking, we can easily generate hypothetical studies $\prob p$ with varying extend of structural  dis\textsc{p}arity between $\mathcal D_X=\{\texttt{1+\,\,pack\,\,daily},\texttt{non-smoker}\}$.  Due to convexity, the violation of symmetry \eqref{eq:StructuralParity:Symmetry} reflected on the two-way  marginal
\equ{
\label{eq:sim:disParity}
p_{X,\boldsymbol S}(x, \texttt{occupation}, \texttt{age}) = \delta \,f_X(x)\cdot f_{\boldsymbol S} (\texttt{occupation}, \texttt{age}) + \left(1-\delta\right) f_{X,\boldsymbol S}(x, \texttt{occupation}, \texttt{age})
}
can be efficiently parametrized by a parameter $\delta\in[0,1]$.
To avoid unintentionally introducing additional bias~\cite{shore1980axiomatic}, the dis\textsc{p}arate table $\prob p=\prob p_{\delta}$  must minimize the $I$-divergence \eqref{eq:KLmetric} from the empirical distribution $\prob f$ of the original study under \eqref{eq:sim:disParity} and confounder \textsc{r}ealism \eqref{eq:Smokers:ConfounderRealism}. This optimization appropriately fixes the association of exposure to disease as well as the three-factor association.
Hence, the simulated contingency table  
described by $\prob p$ (the \textsc{dp}-projection of $\prob f$) will share the same prevalences with the original study. Furthermore, $\prob p$ would by construction result in the same stratified analysis of Table~\ref{tb:SmokeLung:StratifiedOR} corresponding to the same effect size estimate \eqref{eq:Smokers:PR_Intervention_OR} for exposure.

In Figure~\ref{fig:simulations}, we start from the observed dis\textsc{p}arity $\prob f_{X,\boldsymbol S}$ for $\delta=0$ and flow by varying $\delta\rightarrow 1$ towards the structurally symmetric scenario of the  \textsc{pr} projection. 
Concerning the \textsc{mh} ansatz, we recognize two alarming issues. On the one hand, the \textsc{mh} estimator pools specifics from the underlying structural heterogeneity of the simulated studies. Failing to entirely decouple the effect size estimation from  structural variations in the association between $X$ and $\boldsymbol S$, it maps the stratified analysis of Table~\ref{tb:SmokeLung:StratifiedOR} to many different pooled effect size estimates depending on $\delta$.   
%
On the other hand, 
the pooled $\widehat{OR}$ appears to systematically capture something entirely different from the two-way $OR_{\prob q}$ of exposure in \eqref{eq:Smokers:PR_Intervention_OR} corresponding to the confounding-free scenario (dashed line). $\widehat{OR}$ monotonically interpolates  from overestimating the effect size at  the observed dis\textsc{p}arity to underestimating the effect size of exposure at structural \textsc{p}arity, where even the crude estimator $OR_{\prob p_{\delta=1}}$ captures by definition  the confounding-free $OR_{\prob q}$.

\section{An ideally symmetric paradigm}
\label{sc:SymmetricExample}

\begin{figure}[t]
\centering
\raisebox{-0.5\height}{\subfloat[\centering Three-factor graph]{
\begin{tikzpicture}[scale=.92, remember picture ] 
\draw (0,0) circle [radius=0.3] node (Y) {$Y$};
\coordinate (s) at (-1.5,-2);
\draw (s) circle [radius=0.3] node (S) {$S$};
\coordinate (x) at (1.5,-2);
\draw (x) circle [radius=0.3] node (X) {$X$};
%
\draw (0,-2) node[cross=10pt,red,line width=0.4mm,] {};
\draw (-1.1,-0.8) node[rotate=54] {\textsc{r}ealism};
\draw (0,-2.5) node[red] {structural \textsc{p}arity};
\draw (1.2,-0.7) node[rotate=-54,blue] {Intervention};
\draw[-] (S) -- (Y);
\draw[-] (S) -- (X);
\draw[-,dashed,blue] (X) -- (Y);
\end{tikzpicture}\label{graph:three-way}
}
}
\quad~~
\subfloat[\centering Three-way contingency table of ideal study]{\renewcommand{\arraystretch}{1.05}
    \centering
    \begin{tabular}{lll|r}
         $Y$ & $X$ & $S$ & $N\prob f$\\
        \hline &&&\\[-2.5ex]
        \multirow{4}{*}{ \texttt{no\,\,event}} & \multirow{2}{*}{\texttt{control}} & \texttt{A} & 145 \\
         && \texttt{B} & 55\\[0.6ex]
         \rowcolor{\tablecolor}
         & & \texttt{A} & 95\\
         \rowcolor{\tablecolor}
         &\multirow{-2}{*}{\texttt{intervention}}  & \texttt{B} & 5\\
         \hdashline &&&\\[-2ex]
         \multirow{4}{*}{ \texttt{event}} & \multirow{2}{*}{\texttt{control}} & \texttt{A} & 5 \\
         && \texttt{B} & 95\\[0.6ex]
         \rowcolor{\tablecolor}
         &  & \texttt{A} & 55\\
         \rowcolor{\tablecolor}
         &\multirow{-2}{*}{\texttt{intervention}} & \texttt{B} & 145 \\
    \end{tabular} 
    \label{tb:SimpleExample}
}
\caption{Idealized study with two casual factors, $X$ representing the independent variable and $S$ the covariate. By study design there is no confounding.}\label{fig:SimpleExample}
\end{figure}

To illustrate the fundamentally problematic issues with the conventional \textsc{mh} approach, we consider a highly symmetric scenario with complete structural \textsc{p}arity depicted in the graph of Figure~\ref{graph:three-way}. 
For simplicity, we take all one-site marginals to be fully  balanced,
\equ{
\label{eq:SimpleExample:Prevalences}
f_Y(\texttt{no\,\,event}) = f_Y(\texttt{event}) = f_S(A) = f_S(B) = f_X(\texttt{control}) = f_X(\texttt{intervention}) = 0.5
~,
}
so that 
the $X-S$ marginal coincides with the uniform distribution, 
\equ{
\label{eq:SimpleExample:Parity}
f_{X,S}(x,s)=0.25
~.
}
In other words, we have the ideal experimental design where both study groups trivially exhibit identical  structure w.r.t.\ the confounder variable.
Consequently, the \textsc{pr}-projection of $\prob f$ in this case would trivially coincide with $\prob f$ itself.
A realization with $N=600$ participants is given in Table~\ref{tb:SimpleExample}. 

In such symmetric setting, the empirically estimated association  of $Y$ to $X$ is by definition 
the interventional effect size we are after.
All information about this association is contained in the marginal distribution $\prob f_{Y,X}$ after marginalizing over the confounder $S$. 
From  $\prob f_{Y,X}$ we can compute various two-way metrics to assess the strength of association like the Odds ratio
\equ{
\label{eq:SimpleExample:Intervention_OR}
OR_{\prob f} = \frac{Odds_{\prob f}(\texttt{event}\vert \texttt{intervention})}{Odds_{\prob f}(\texttt{event}\vert \texttt{control})} = 4
~,
}
where
\equ{
Odds_{\prob f}(\texttt{event}\vert x) = \frac{f_{Y,X}(\texttt{event}, x)}{f_{Y,X}(\texttt{no\,\,event}, x)} \quads{\text{for}} x\in\{\texttt{control},\texttt{intervention}\}
~.
}
Conversely, we may look at  the risk ratio
\equ{
\label{eq:SimpleExample:Intervention_RR}
RR_{\prob f} = \frac{f_{Y\vert X}(\texttt{event}\vert\texttt{intervention})}{f_{Y\vert X}(\texttt{event}\vert\texttt{control})} = 2
}
expressed in terms of empirical conditional
\equ{
f_{Y\vert X}(\texttt{event}\vert x) = \frac{f_{Y, X}(\texttt{event}, x)}{f_{X}(x)}
~.
}
Despite that the two metrics vary in their numerical values, they describe the exact same, by construction confounding-free situation, as they descend from the same three-way contingency table~\ref{fig:SimpleExample}.

As far as the effect size estimate in this ideal experimental setting is concerned, we do not need to compute anything beyond \eqref{eq:SimpleExample:Intervention_OR} or \eqref{eq:SimpleExample:Intervention_RR}. 
Of course, no-one prevents us from calculating three-way odds ratios or relative risks from Table~\ref{fig:SimpleExample}, stratified by $S$:
\equ{
\label{eq:SimpleExample:Stratified_OR}
\frac{Odds_{\prob f}(\texttt{event}\vert \texttt{intervention}, \texttt A) }{Odds_{\prob f}(\texttt{event}\vert \texttt{control}, \texttt A)}
=
\frac{Odds_{\prob f}(\texttt{event}\vert \texttt{intervention}, \texttt B) }{Odds_{\prob f}(\texttt{event}\vert \texttt{control}, \texttt B)}
=
\frac{319}{19} \approx 16.79
}
and
\equ{
\label{eq:SimpleExample:Stratified_RR}
\frac{f_{Y\vert X,S}(\texttt{event}\vert \texttt{intervention}, \texttt A)}{f_{Y\vert X,S}(\texttt{event}\vert \texttt{control}, \texttt A)}
= 11
\quads{,}
\frac{f_{Y\vert X,S}(\texttt{event}\vert \texttt{intervention},\texttt  B)}{f_{Y\vert X,S}(\texttt{event}\vert \texttt{control}, \texttt B)}
= 
\frac{29}{19} \approx 1.53~,
}
where as before
\equ{
f_{Y\vert X,S}(\texttt{event}\vert x,s) = \frac{f(\texttt{event},x,s)}{f_{X,S}(x,s)} \quads{\text{for}} s\in\{\texttt{A},\texttt{B}\}~.
}
Out of these three-way metrics, \textsc{mh} pools 
\equ{
\label{eq:SimpleExample:PooledEstimators}
\widehat{OR} \approx 16.79
\qand
\widehat{RR} \approx 3.17
~.
}

Being a weighted average~\cite{NomaNagashima_2016_19_35}, the first pooled estimate coincides with stratified odds \eqref{eq:SimpleExample:Stratified_OR}, which happen to be equal in this symmetric scenario. 
In contrast, the pooled relative risk cannot be motivated at all within the three-way risk metrics \eqref{eq:SimpleExample:Stratified_RR}.
Clearly, these \textsc{mh} metrics fail to capture the effect size estimate of the two-way $OR$ and $RR$ in \eqref{eq:SimpleExample:Intervention_OR} and \eqref{eq:SimpleExample:Intervention_RR} respectively.
Most alarmingly, there exists no three-way contingency table with the given prevalence rates \eqref{eq:SimpleExample:Prevalences} and structural \textsc{p}arity \eqref{eq:SimpleExample:Parity} that can accommodate both pooled estimates \eqref{eq:SimpleExample:PooledEstimators} with non-negative cell counts. 
Therefore, $\widehat{OR}$ and $\widehat{RR}$ cannot be formally thought of as odds ratios or relative risks in a relevant confounding-free scenario.

\paragraph{}
This raises a fundamental question of both theoretical and practical importance:
\textit{what precisely do such pooled estimators capture and what does the corresponding statistical testing aim to assess?}
From our findings, we can certainly say what they do not capture: \textit{the two-way, confounding-free effect size estimate.}
This highly symmetric example illustrates the importance of the top-down approach in flow diagram~\ref{fig:Top-Down}. In the worst case, declared two-way metrics that have been motivated in a bottom-up fashion might fail to be lifted to a three-way contingency table, i.e.\ they lack any epistemological grounds for consistent interpretation.



\subsection*{Code and data availability}
An \textsc{r}-package implementing the \textsc{p}(\textsc{u})\textsc{r} projection of empirical distribution $\prob f$ (or of uniform distribution $\prob u$) introduced in~\cite{loukas2023demographic}, alongside the historic datasets used in~\cite{loukas2023eliminating} and the present paper, are available directly from the authors upon reasonable request.

\renewcommand*{\bibfont}{\small}
\printbibliography

\end{document}